**A GPU-based Solver for Polarization Dynamics in Ferroelectric Materials**

Ali Hasan[1], Edoardo Piccolo[2], Anna Giordano[3], Natalya Fedorova[4], Jorge Íñiguez-González[4,5], Davi Rodrigues[2], Giovanni Finocchio[1]

[1]Department of Mathematical and Computer Sciences, Physical Sciences, and Earth Sciences, University of Messina, I-98166, Messina, Italy
[2]Department of Electrical and Information Engineering, Politecnico di Bari, I-70125 Bari, Italy
[3]Department of Engineering, University of Messina, I-98166, Messina, Italy
[4]Smart Materials Unit, Luxembourg Institute of Science and Technology (LIST), 5 avenue des Hauts-Fourneaux, L-4362 Esch/Alzette, Luxembourg
[5]Department of Physics and Materials Science, University of Luxembourg, Rue du Brill 41, L-4422 Belvaux, Luxembourg

**Abstract:**

Ferroelectric materials can be used for the development of multiple device concepts combining non-volatility, small dimensions, low-power actuation, and electrical tunability. Such development demands efficient and precise design of simulation tools describing the polarization texture. However, most existing ferroelectric solvers are CPU-based and rely on simplified electrostatic treatments and reduced-dimensional representations of the polarization field. These approximations limit their ability to capture finite-size and boundary effects and restrict the range of domain structures and domain walls that can be realistically simulated. Here, we present a fully GPU (graphics processing units)-accelerated and scalable numerical solver, named PETASPIN_microelectrics, for computing the full polarization vector field of ferroelectric systems using the Ginzburg–Landau formalism. Our solver incorporates an optimized and validated calculation of the full electrostatic field and enables the parallel execution of multiple simulations. We systematically validated the solver with several benchmark problems, including phase transitions in $BaTiO_3$ and ferroelectric domain wall profiles. Our simulations reproduce temperature-driven hysteretic phase transitions in $BaTiO_3$. We also reproduce hysteresis loops and demonstrate stabilization of a three-dimensional hybrid skyrmion in a $PbTiO_3/SrTiO_3$ bilayer system. Our results show quantitative agreement with predictions from an analytical theory and prior experimental studies. The proposed solver provides an efficient, accurate platform for large-scale simulations of ferroelectric materials including stabilization of topological textures supporting predictive modeling for next-generation of ferroelectric device design.

## 1. Introduction

Ferroelectric materials have long attracted considerable interest both from fundamental and applied viewpoints [1–3], [4–6], and the interest continues to be fueled e.g. with the recent discovery of room temperature nanoscale topological textures [4,6,7]. Such rich behaviour has the potential to be used in applications including non-volatile memories, neuromorphic computing architectures, and energy-storage devices [8–11]. Polarization can be manipulated by external stimuli including electric fields [3,12], mechanical strain [3,12], and thermal excitations [7]. In addition, the ability to manipulate and reconfigure nanoscale domain textures enables new opportunities for the application of ferroelectric materials [13–17]. As the field continues to grow and the emphasis shifts toward nanoscale textures, more accurate modeling tools capable of resolving ferroelectric domain structures have become increasingly essential. Phase-field simulations based on the time-dependent Ginzburg–Landau (TDGL) equation constitute one of the most powerful approaches for investigating such domain structures in ferroelectrics at equilibrium [18–20]. This simulation approach is based on the energy minimization involving different contributions such as Landau free energy, gradient and electrostatic energies [21]. The Landau free energy is a local contribution whereas gradient energy accounts for the coupling between neighboring regions in space [22]. However, the quantitative behaviour, and in some cases even the qualitative response especially for studying texture stability, of ferroelectrics depend on the numerical approach used to evaluate the electrostatic field, which is a computationally expensive nonlocal long-range interaction [23–26]. Although CPU-based phase-field simulations in Fourier space enable the modeling of 2-dimentional and 3-dimentional ferroelectric domain structures, the long-range nature of electrostatic interactions make such simulations computationally demanding for large samples, long relaxation dynamics, or to perform extensive systematic studies [24,27]. For such reason, many existing ferroelectric solvers rely on simplified electrostatic approximations or reduced-dimensional schemes [23,28], and there is a need to develop high performance computational tools capable of accurately resolving full three-dimensional ferroelectric dynamics. To address such challenge, GPU-accelerated frameworks such as FerroX have been developed to efficiently scale 3-dimentional phase-field ferroelectric simulations on many-core hardware; however, their current single-component (uniaxial) polarization formulation limits the description of polarization rotation, vectorial coupling, and complex three-dimensional domain and topological structures [29].

In this work, we introduce PETASPIN_microelectrics, a fully GPU-accelerated and scalable solver for modeling ferroelectric polarization phase stability covering all major energy contributions in the phase-field framework and having a highly optimized electrostatic field calculation. The solver explicitly resolves the full three-dimensional polarization vector $\boldsymbol{P} = (P_x, P_y, P_z)$ in both space and time overcoming the single-component polarization limitations of existing GPU-accelerated phase-field frameworks. Built natively on CUDA™ (Compute Unified Device Architecture) C/C++ and the Thrust library [30,31], PETASPIN_microelectrics achieves high computational efficiency while remaining easily extensible. The solver supports spatially varying material parameters, time-dependent excitations, multiple dynamical integration schemes, and parallel execution of independent simulations, which enables rapid exploration of material responses across a wide range of conditions.

We validated PETASPIN_microelectrics by reproducing the experimentally observed sequence of temperature-driven phase transitions in $BaTiO_3$. Furthermore, we recovered the analytical solution for domain wall profiles in presence of gradient and Landau free energies, with excellent quantitative agreement. We have also studied field-driven hysteresis and skyrmion stabilization in $PbTiO_3/SrTiO_3$ bilayers. Manipulation of these domain patterns are critical both for fundamental understanding and for exploring design of next-generation devices [32]. Overall, our results demonstrate the accuracy and versatility of PETASPIN_microelectrics, establishing it as a predictive tool for ferroelectric research and device design.

## 2. Ferroelectric Model

As commonly done, we assume that the relaxation processes of the ferroelectric order parameter, the polarization vector $\boldsymbol{P}(\boldsymbol{r}, t)$, are well-described by the time-dependent Ginzburg–Landau (TDGL) equation [1]:

$$\frac{dP_i(\boldsymbol{r},t)}{dt} = -L\frac{\delta F}{\delta P_i(\boldsymbol{r},t)} \tag{1}$$

Here, $i = x, y, z$, and $L$ represents the coefficient controlling the time scale for thermalization, besides affecting kinetic properties such as domain-wall mobility [33]. In all simulations, for simplicity, we set $L = 1.0\ \mathrm{s^{-1}m^{-2}C^{2}N^{-1}}$; however, the physically accurate value of $L$ should be determined from experimental fitting or predictive atomistic simulations on a case-by-case basis Ref. [34]. The total free energy $F$ includes the Landau free energy ($F_{Land}$), the gradient (Ginzburg) energy ($F_{Grad}$), the electrostatic energy ($F_{Elec}$), and the electrical energy density associated with an external electric field ($F_{Ext}$).

$$F = F_{Land} + F_{Grad} + F_{Elec} + F_{Ext} \tag{2}$$

The Landau free energy ($F_{Land}$) can be viewed as a Taylor expansion on the polarization, around a paraelectric reference phase, up to an appropriate order, with odd powers omitted due to symmetry considerations [11]. For simplicity and concreteness, in the following we discuss the prototypical case of ferroelectric perovskite oxides like $BaTiO_3$ or $PbTiO_3$, generalization to other ferroelectric families being straightforward. Below the transition temperature, this energy is minimized by the development of a non-zero polarization magnitude, which enables the description of different ferroelectric phases as a function of temperature [35]. In this work, we employ a Landau free energy expansion up to sixth-order terms as in [25] for the three-dimensional polarization vector $\boldsymbol{P} = (P_x, P_y, P_z)$, expressed as follows:

$$F_{Land} = \alpha_0(T - T_0)(P_x^2 + P_y^2 + P_z^2) + \alpha_{11}(P_x^4 + P_y^4 + P_z^4) + \alpha_{12}(P_x^2P_y^2 + P_y^2P_z^2 + P_z^2P_x^2) + \alpha_{111}(P_x^6 + P_y^6 + P_z^6) + \alpha_{112}\left(P_x^4(P_y^2 + P_z^2) + P_y^4(P_x^2 + P_z^2) + P_z^4(P_x^2 + P_y^2)\right) + \alpha_{123}(P_x^2P_y^2P_z^2) \tag{3}$$

Here $T$ denotes the temperature, while $T_0$ is the temperature at which the paraelectric phase loses its stability; and $\alpha_i$, $\alpha_{ij}$ and $\alpha_{ijk}$ are dielectric stiffnesses at different orders in the expansion [36].

The gradient (Ginzburg) energy adopted from [25] accounts for spatial variations in polarization, particularly important in systems with domain structures where regions of uniform polarization are separated by domain walls. It is expressed as:

$$F_{Grad} = \frac{1}{2}G_{11}(P_{x,x}^2 + P_{y,y}^2 + P_{z,z}^2) + G_{12}(P_{x,x}P_{y,y} + P_{y,y}P_{z,z} + P_{z,z}P_{x,x}) + \frac{1}{2}G_{44}\left((P_{x,y} + P_{y,x})^2 + (P_{y,z} + P_{z,y})^2 + (P_{x,z} + P_{z,x})^2\right) + \frac{1}{2}G'_{44}\left((P_{x,y} - P_{y,x})^2 + (P_{y,z} - P_{z,y})^2 + (P_{x,z} - P_{z,x})^2\right) \tag{4}$$

where $P_{i,j} = \frac{\partial P_i}{\partial x_j}$ and $G_{ij}$ are the gradient energy coefficients.

We note that the expressions for the Landau and gradient free energies depend on the specific material properties and crystal symmetries [29,37–39]. Additional terms can be incorporated to describe more complex materials or alternative symmetry classes as required.

The electrostatic energy originates from long-range interactions among dipoles, which favour configurations where dipoles align head-to-tail [25]**.** The full form of the electrostatic energy employed in this work is expressed as

$$F_{Elec} = -\frac{1}{2}\int_V \boldsymbol{E}_{Elec}(\boldsymbol{r}).\boldsymbol{P}(\boldsymbol{r})\, d^3r \tag{5}$$

where $\boldsymbol{E}_{Elec}$ is the electrostatic field generated by the bound charges associated with the polarization field. The electrostatic field is computed from the volume and surface bound charge densities, $\rho_V = -\nabla \cdot \boldsymbol{P}$ and $\sigma_S = \boldsymbol{P} \cdot \boldsymbol{n}$ , through the standard Coulomb kernel [40,41],

$$\boldsymbol{E}_{Elec}(\boldsymbol{r}) = \frac{1}{4\pi\epsilon_0\epsilon_B}\left[\int_{V'} \frac{(\boldsymbol{r} - \boldsymbol{r}')\rho_{V(\boldsymbol{r}')}}{|\boldsymbol{r} - \boldsymbol{r}'|^3} dr'^3 + \int_{S'} \frac{(\boldsymbol{r} - \boldsymbol{r}')\sigma_{S(\boldsymbol{r}')}}{|\boldsymbol{r} - \boldsymbol{r}'|^3} dr'^2\right] \tag{6}$$

where $\epsilon_0$ is the vacuum permittivity and $\epsilon_B$ is the background permittivity which quantifies the electronic screening. The treatment of the electrostatic field differs among available phase-field solvers, particularly with respect to the imposed boundary conditions, geometric constraints, system size, and the choice of background permittivity $\epsilon_B$ [1,42–44]. In this work, we consider finite samples embedded in vacuum, with the polarization set to $\boldsymbol{P} = 0$ outside the sample. Since no standardized benchmark currently exists for electrostatic field solvers in ferroelectric systems, we validated our implementation against established approaches used for magnetostatic field calculations [40]. Given the variability across different numerical implementations, we treat $\epsilon_B$ as an effective fitting parameter to ensure consistency with experimentally relevant conditions that can be set freely in our simulation tool.

The external electrical energy describes how an electric field $\boldsymbol{E}_{Ext}$ influences the polarization within the material. This external field tends to align the dipoles along its direction and modify the domain structure [45],

$$F_{Ext} = -\int \boldsymbol{E}_{Ext}.\boldsymbol{P}(\boldsymbol{r})\, d^3r \tag{7}$$

For the numerical simulations presented in this work, we used material parameters taken from the literature, as summarized in Table 1.

Table 1. Material parameters used in the simulations for $BaTiO_3$ [46,47], $PbTiO_3$ [48] and $SrTiO_3$ [49].

| Parameter | Unit | $BaTiO_3$ | $PbTiO_3$ | $SrTiO_3$ |
|---|---|---|---|---|
| $\alpha_1 = \alpha_0(T - T_0)$ | $Nm^2C^{-2}$ | $3.34 \times 10^5(T - 381)$ | $3.85 \times 10^5(T - 752)$ | $405[\coth(54/T) - \coth(54/30)]$ |
| $\alpha_{11}$ | $Nm^6C^{-4}$ | $4.69 \times 10^6(T - 393) - 2.02 \times 10^8$ | $-7.3 \times 10^7$ | $1.7 \times 10^9$ |
| $\alpha_{12}$ | $Nm^6C^{-4}$ | $3.23 \times 10^8$ | $7.5 \times 10^8$ | $1.37 \times 10^9$ |
| $\alpha_{111}$ | $Nm^{10}C^{-6}$ | $-5.52 \times 10^7(T - 393) + 2.76 \times 10^9$ | $2.6 \times 10^8$ | – |
| $\alpha_{112}$ | $Nm^{10}C^{-6}$ | $4.47 \times 10^9$ | $6.1 \times 10^8$ | – |
| $\alpha_{123}$ | $Nm^{10}C^{-6}$ | $4.91 \times 10^9$ | $-3.7 \times 10^9$ | – |
| $G_{11}$ | $Nm^4C^{-2}$ | $5.1 \times 10^{-10}$ | $1.038 \times 10^{-10}$ | – |
| $G_{12}$ | $Nm^4C^{-2}$ | $-2.0 \times 10^{-11}$ | 0.0 | – |
| $G_{44}$ | $Nm^4C^{-2}$ | $2.0 \times 10^{-11}$ | $5.19 \times 10^{-11}$ | – |
| $G'_{44}$ | $Nm^4C^{-2}$ | $2.0 \times 10^{-11}$ | $5.19 \times 10^{-11}$ | – |
| $G_{110}$ | $Nm^4C^{-2}$ | $5.1 \times 10^{-10}$ | $1.73 \times 10^{-10}$ | – |

To improve numerical stability and avoid overflow, the computations are performed using dimensionless parameters without loss of generality. Table 2 summarizes the normalization factors used to obtain the corresponding dimensionless quantities.

Table 2. Definitions of the normalized parameters used in the numerical simulations.

| Parameter | Normalized Parameter | Parameter | Normalized Parameter | Parameter | Normalized Parameter |
|---|---|---|---|---|---|
| $P_i$ | $P_i^* = \frac{P_i}{P_0}$ | $\alpha_{111}$ | $\alpha_{111}^* = \frac{\alpha_{111} P_0^4}{\lvert\alpha_1\rvert}$ | $G_{12}$ | $G_{12}^* = \frac{G_{12}}{G_{110}}$ |

| $t$ | $\tau = \|\alpha_1\| L t$ | $\alpha_{112}$ | $\alpha_{112}^* = \frac{\alpha_{112} P_0^4}{\|\alpha_1\|}$ | $G_{44}$ | $G_{44}^* = \frac{G_{44}}{G_{110}}$ |
|---|---|---|---|---|---|
| $\alpha_1$ | $\alpha_1^* = \frac{\alpha_1}{\|\alpha_1\|}$ | $\alpha_{123}$ | $\alpha_{123}^* = \frac{\alpha_{123} P_0^4}{\|\alpha_1\|}$ | $G'_{44}$ | $G'^*_{44} = \frac{G'_{44}}{G_{110}}$ |
| $\alpha_{11}$ | $\alpha_{11}^* = \frac{\alpha_{11} P_0^2}{\|\alpha_1\|}$ | $r_i$ | $r_i^* = \sqrt{\frac{\|\alpha_1\|}{G_{110}}} r_i$ | $E_{Ext}$ | $E_{Ext}^* = \frac{E_{Ext}}{\|\alpha_1\| P_0}$ |
| $\alpha_{12}$ | $\alpha_{12}^* = \frac{\alpha_{12} P_0^2}{\|\alpha_1\|}$ | $G_{11}$ | $G_{11}^* = \frac{G_{11}}{G_{110}}$ | – | – |

## 3. Technical details of the microelectric solver

The GPU-native PETASPIN_microelectrics solver is developed to numerically solve the TDGL equation using a finite-difference approach. The polarization vector $\boldsymbol{P}(\boldsymbol{r}, t)$ is assumed to be uniform within each computational cell of the finite-difference grid. Based on the gradient energy parameters, the dimensions of each cell can be chosen small enough so that the polarization vector varies continuously from cell to cell in regions between domains of different orientations. The characteristic length scale depends on several factors, including the Landau free energy, gradient energy and electrostatic energy. The importance of these contributions can vary depending on the size of the structure under investigation, so the grid resolution must be adapted to the specific physical phenomena being simulated [35,50,51].

To achieve high computational performance, the solver leverages both CUDA™ C/C++ and the Thrust library. The dominant computational cost arises from evaluating the electrostatic field, which involves long-range non-local interactions. This component is computed using NVIDIA's CUDA™ Fast Fourier Transform (FFT) library, complemented by custom CUDA™ kernels for intermediate operations, similar to what is used for magnetostatics [40]. The contributions from gradient energy, Landau free energy, and external-field energy are computed using the Thrust library. Thrust abstracts many low-level aspects of CUDA™ kernel development, its iterator-based structure eliminates common sources of memory-access errors, while automated kernel launch optimization achieves performance comparable to custom optimized CUDA™ kernels. These features in PETASPIN_microelectrics solver reduce code complexity, enhance robustness, and facilitate rapid integration of other physical contributions, making it flexible and easily extensible.

The current version of PETASPIN_microelectrics supports spatially varying material parameters at the cell level and time-dependent external excitations. A further distinguishing feature compared with existing microelectric solvers is its ability to run multiple simulations in parallel on the GPU, which provides efficient parameter sweeps and studies involving external fields, temperature variation, or material non-uniformity. Running multiple simulations in parallel also reduces the overall computational cost significantly compared with executing each simulation sequentially. It includes as time integration schemes the Heun's method,

Runge–Kutta (RK45) [52], a fourth-order method with adaptive time stepping, and Adams–Bashforth (AB3M2) [53], a multistep explicit scheme.

All input parameters are kept in a configuration folder, where flags allow users to turn different contributions on or off. The user has the option to normalize these input parameters to obtain dimensionless quantities, as listed in Table 2. The solver records cell-level and averaged values of polarization, fields, and energy at the intervals defined by the user. PETASPIN_microelectrics overcomes the intrinsic limitations of FerroX allowing to study polarization rotation, topological textures such as vortices and skyrmions, and phase transitions.

## 4. Results and Discussion

### 4.1. Polar domains and domain wall formation

Within the TDGL framework, the combined Landau free energy and gradient contributions drive the formation of polar domains in which the polarization field is spatially uniform. These domains correspond to ground states of the bulk system. The Landau free energy determines a temperature-dependent stable finite polarization magnitude along specific crystallographic axes. However, it does not select a unique orientation, allowing for the emergence of domains. In the absence of gradient terms, domain walls would have zero width. The inclusion of gradient energy penalizes sharp spatial variations of the polarization, leading to domain walls with a finite characteristic width. In general, both the domain size and the domain-wall width at zero external field are governed by the interplay among the Landau free, the gradient, and the electrostatic energies.

We validated the implementation of the field arising from the gradient and Landau free energies by comparing it with analytical solutions using a simplified one-dimensional energy model. In this model, for simplicity, we set $G_{12} = G_{44} = G'_{44} = 0$ and retained only a non-zero value of $G_{11}$, considering a single component of the polarization vector. The resulting energy expression is

$$E(P) = \int \left[ \alpha_1 P^2 + \alpha_{11} P^4 + \frac{1}{2} G_{11} \left( \frac{dP}{dx} \right)^2 \right] dx \tag{8}$$

The first two terms arise from the Landau free energy expansion, while the last term corresponds to the gradient energy. By applying the Euler–Lagrange equation, we obtain the analytical solution for the domain-wall profile, which follows a hyperbolic tangent form [54,55]:

$$P(x) = \sqrt{\frac{|\alpha_1|}{2\alpha_{11}}} \tanh\left( \frac{(x - x_0)}{\sqrt{G_{11}/|\alpha_1|}} \right) \tag{9}$$

where $x_0$ denotes the center of the domain wall. This solution is valid for domain walls with a single non-vanishing polarization component modulated along the $x$-direction. Such domain walls, however, are energetically unfavourable in the presence of electrostatic interactions due to charge accumulation at the wall. For domain walls oriented perpendicular to the x-direction, an analogous expression can be obtained, with the gradient coefficient $G_{11}$ replaced by the effective coefficient $G_{44} + G'_{44}$, and is given by

$$P(x) = \sqrt{\frac{|\alpha_1|}{2\alpha_{11}}} \tanh\left(\frac{(x - x_0)}{\sqrt{(G_{44} + G'_{44})/|\alpha_1|}}\right) \tag{10}$$

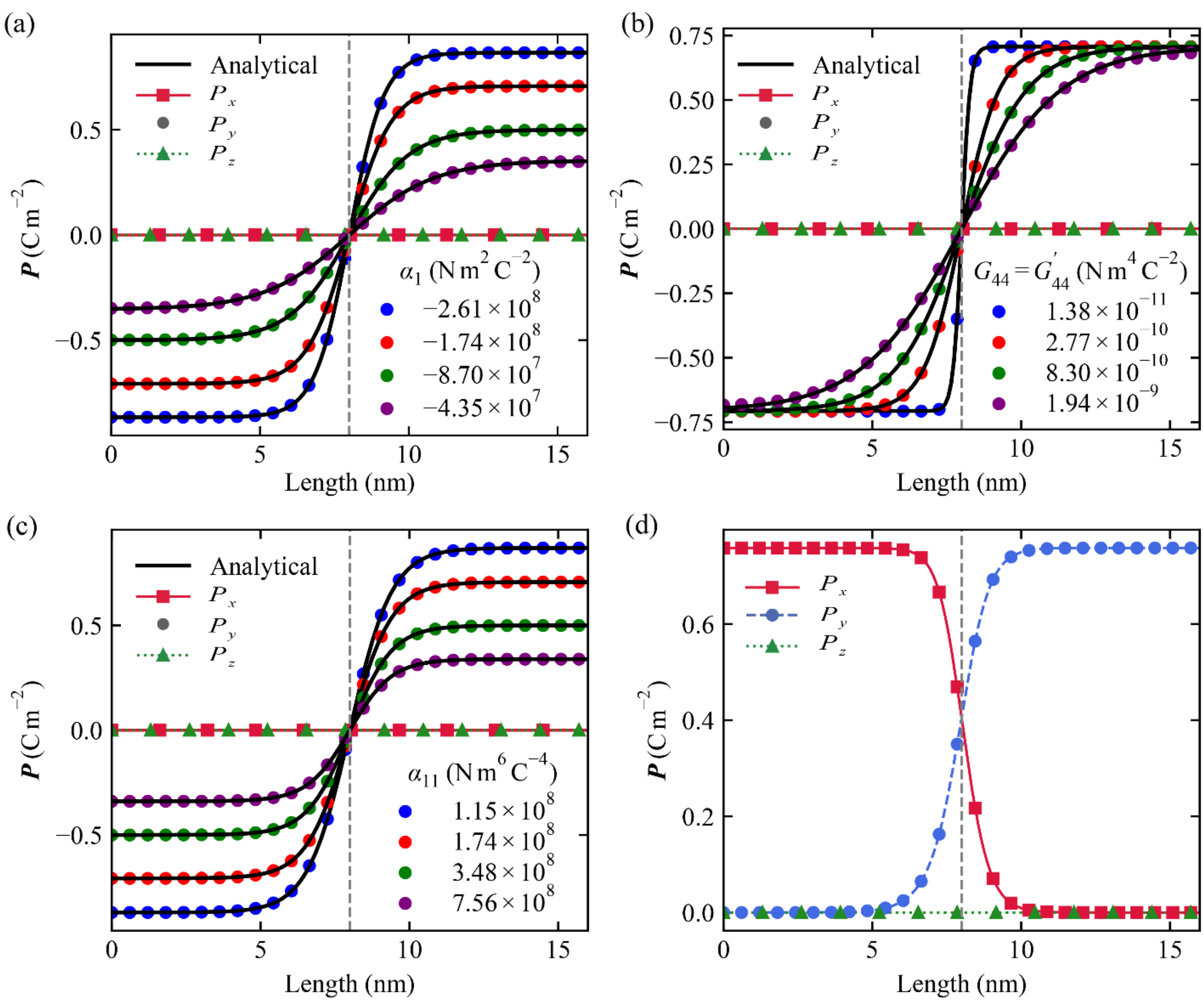

Figure 1. Polarization profiles across ferroelectric domain walls and the influence of Landau and gradient energy parameters. (a) Polarization profiles of 180° domain walls along the wall-normal direction, illustrating the effect of varying the Landau coefficient $\alpha_1$. (b) Effect of varying the gradient constants $G_{44} = G'_{44}$ on the 180° wall profile. (c) Influence of the higher-order Landau coefficient $\alpha_{11}$ on the 180° domain-wall. Solid

black lines denote analytical solution, while symbols represent numerical results. (d) Polarization profile of a 90° domain wall.

To examine the impact of the gradient field and to validate our numerical implementation we have simulated a one-dimensional grid. Unless otherwise specified, for the 180° domain-wall profiles, the Landau coefficients were set to $\alpha_1 = -1.7402 \times 10^8$ Nm$^2$C$^{-2}$ and $\alpha_{11} = 1.7402 \times 10^8$ Nm$^6$C$^{-4}$, while the gradient energy coefficients were chosen as $G_{44} = G'_{44} = 1.73 \times 10^{-10}$ Nm$^4$C$^{-2}$ with $G_{11} = G_{12} = 0$. For the 90° domain-wall calculations, the Landau coefficients were taken from Table 1, and the gradient energy coefficients were fixed to $G_{11} = G_{44} = G'_{44} = 1.73 \times 10^{-10}$ Nm$^4$C$^{-2}$ with $G_{12} = 0$. Fig.1(a) shows the agreement between analytical and numerical polarization profile of 180° domain wall and impact of varying $\alpha_1$. The domain wall profile is characterized by a change in the amplitude of the polarization rather than by a rotation of the polarization direction. The polarization magnitude gradually decreases from the lefthand side, vanishes at the domain-wall center, and then increases toward the righthand side. Increasing $\alpha_1$ in magnitude enhances the polarization magnitude and leads to a reduction in the domain-wall width. The gradient energy parameters $G_{44}$ and $G'_{44}$ directly determine the width of the domain wall, as illustrated in Fig. 1 (b), showing excellent agreement with analytical predictions. A similar dependence of the domain-wall width on $G_{11}$ can also be established. Furthermore, increasing the fourth-order Landau free energy coefficient $\alpha_{11}$ reduces the polarization magnitude, as demonstrated in Fig. 1 (c). Fig. 1(d) shows the profile of a 90° domain wall stabilized for $G_{11}$, $G_{44}$, and $G'_{44}$ non-zero values, using Landau free parameters for $PbTiO_3$ at 300 K. In this case, we observe a rotation of the polarization field, corresponding to a smooth transition from non-zero $P_x$ to non-zero $P_y$. For 90° domain wall we found no analytical solution. The analytical solution for the 90° domain wall can, in principle, be obtained by minimizing the following energy functional

$$\begin{aligned} E(P) = \iint \Big( & \alpha_1\left(P_x^2 + P_y^2\right) + \alpha_{11}\left(P_x^4 + P_y^4\right) + \alpha_{12}\left(P_x^2 P_y^2\right) + \alpha_{111}\left(P_x^6 + P_y^6\right) \\ & + \alpha_{112}\left(P_x^4 P_y^2 + P_y^4 P_x^2\right) \\ & + \frac{1}{2}\left(G_{11}\left(P_{x,x}^2 + P_{y,y}^2\right) + G_{44}\left(P_{x,y} + P_{y,x}\right)^2\right. \\ & \left. + G'_{44}\left(P_{x,y} - P_{y,x}\right)^2\right)\Big)\, dxdy \end{aligned} \tag{11}$$

### 4.2. Phases transition in $BaTiO_3$

The phase transitions of $BaTiO_3$ [56,57] are attracting renewed attention and have been revisited in recent studies, including those employing machine-learning enhanced models for $BaTiO_3$ [58] and molecular-dynamics-based simulations [59], which highlight the abrupt jumps in polarization across the transformations.

By setting the time derivative of Eq. 1 equal to zero,

$$\frac{\delta F(P_i)}{\delta P_i} = 0, \tag{12}$$

we can compute the equilibrium states of the polarization. Considering only the $F_{Land}$ introduced in Eq. 3, it is possible to find analytical solutions of equation (12). Three phases are possible, tetragonal with polarization along a ⟨100⟩ direction, orthorhombic ⟨110⟩, and rhombohedral ⟨111⟩, which amount to 26 possible orientations of the polarization. The polarization magnitudes and energies associated with different orientations corresponding to the same symmetry are identical, whereas they differ across the three phases. Figure 2(a) and Figure 2(b) show such analytical solutions (solid lines) for the polarization magnitude and the corresponding Landau free energies, as a function of the temperature (0–360 K) for $BaTiO_3$ parameters from [46], as listed in Table 1.

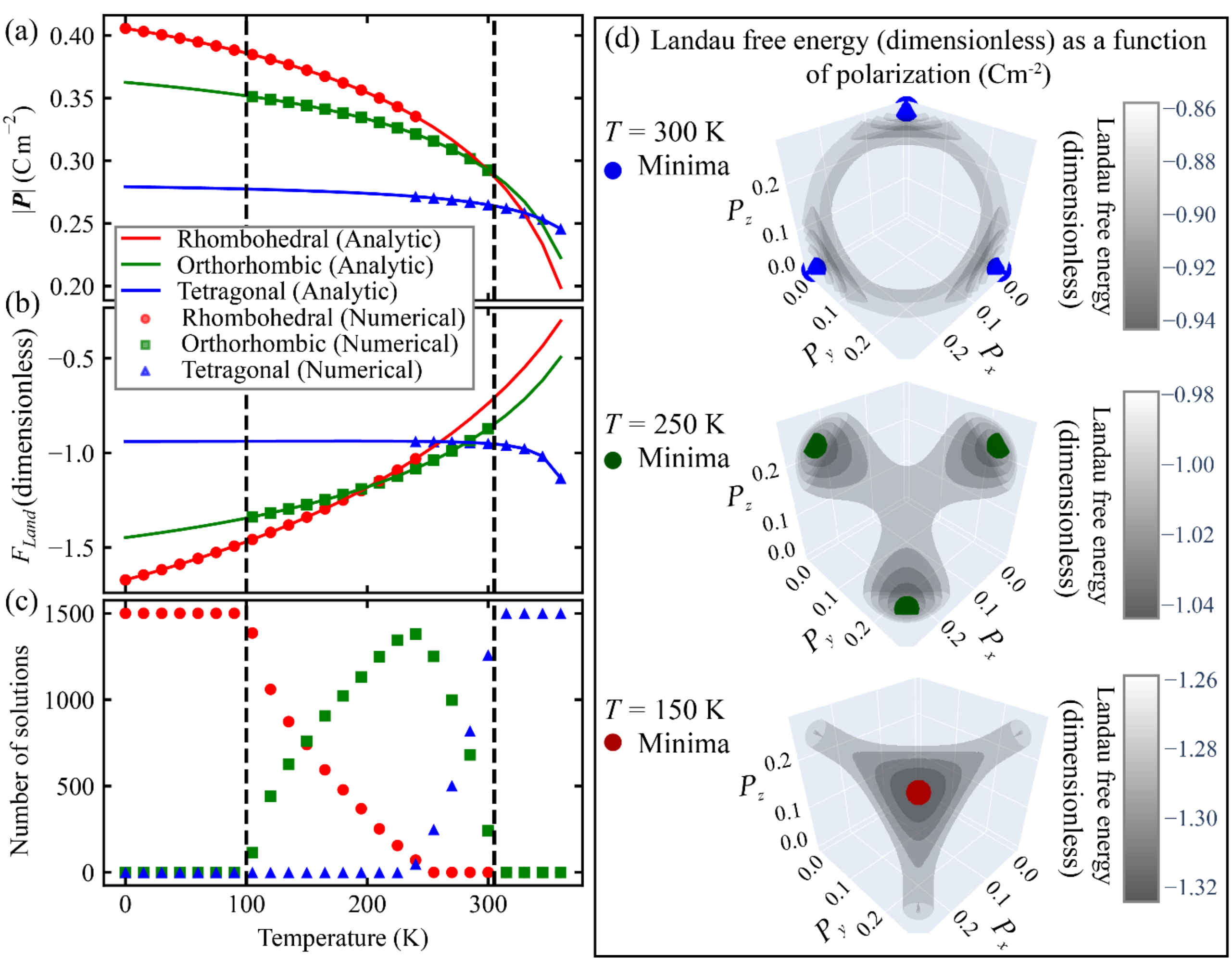

Figure 2. Temperature dependence of (a) polarization magnitude $P$ and (b) Landau free energy $F_{Land}$ and (c) number of solutions in each phase. Panel (d) presents three-dimensional visualizations of the global minima of the Landau free energy surface at representative temperatures (300 K, 250 K, and 150 K), evaluated over the first-octant polarization space ranging from (0.0, 0.0, 0.0) to (0.27, 0.27, 0.27) C m⁻². Gray surfaces indicate isoenergetic contours.

The numerical solutions including only the Landau free energy at a specific cell in the finite-difference grid will always converge to one of the 26 states that coincides with one of the analytical solutions (see comparison in Fig. 2a and 2b).

As can be seen, there are regions of temperatures where two phases co-exist, one being metastable. Thus, for a given temperature, we have performed the calculation of the equilibrium state for 1500 simulations each of them starting from random polarization values with random orientations and magnitudes on the order of $10^{-3}\ \mathrm{Cm}^{-2}$.

Figure 2(c) summarizes the results of those simulations, within 0-100 K, all equilibrium states converge to the rhombohedral phase, above 100 K, the number of solutions corresponding to the rhombohedral phase gradually decreases while the orthorhombic phase begins to appear, accompanied by a reduction in polarization magnitude. With further temperature increase, the orthorhombic phase becomes dominant. At approximately 240 K, the orthorhombic phase reaches its maximum probability, the number of rhombohedral solutions approaches zero, and the tetragonal phase begins to emerge. Above 300 K, the tetragonal phase becomes the dominant state, with all 1500 simulations stabilizing in the tetragonal phase, while the number of orthorhombic solutions found by our procedure drops to zero.

Figure 2(d) represents a three-dimensional visualization of the Landau free energy landscape corresponding to the polarization in the first octant of a 3-dimentional $BaTiO_3$ cube at three temperatures: 300 K, 250 K, and 150 K. For clarity, regions of higher energy are shown as transparent. At 300 K, the energy landscape contains three global minima with tetragonal symmetry, marked in blue, each with an energy of -0.94 (dimensionless). At this temperature, the energy of the orthorhombic phase is also very close to that of the tetragonal phase. The full eight-octant cube contains six minima in total for the tetragonal phase. At 250 K, the global minima shift to the orthorhombic phase, shown with red markers, while the rhombohedral minima appear close in energy; the full cube contains twelve orthorhombic minima. At 150 K, the global minima correspond to the rhombohedral phase, although the orthorhombic minima remain close in energy. The full eight-octant cube contains eight rhombohedral minima in total. The results discussed above are obtained by considering only the Landau free-energy contribution, which captures the intrinsic thermodynamic stability of the homogeneous polarization states and the associated phase transitions. While this approach accurately reproduces the analytical equilibrium states and their temperature evolution, it does not account for spatial polarization variations or long-range electrostatic interactions.

### 4.3. Size dependence polarization in $PbTiO_3$ systems

Electrostatic interactions play a fundamental role in determining the stability of polar configurations, particularly in finite systems where surface effects and background screening strongly modify the energy landscape. In confined geometries, depolarization fields significantly influence the polarization at interfaces and can renormalize phase transition behavior [60–62]. Many phase-field studies adopt simplified electrostatic treatments, such as periodic boundary conditions, approximate background permittivity values, or imposed symmetry constraints [20,63]. Although computationally less expensive, these approximations may fail to capture the full complexity of polarization textures in realistic finite samples, including flux-closure domains, vortices, skyrmions, domain walls, and more intricate three-dimensional structures [64–67].

To validate the capability of our solver to capture finite-size electrostatic effects, we systematically investigated the polarization configurations in $PbTiO_3$ systems as a function of system size. Figure 3 presents the relaxed polarization patterns obtained from random initial configurations for different cube dimensions. Figures 3(b)-(d) show the polarization configurations in the cross section defined in Fig. 3(a) for cubic samples ranging from $5 \times 5 \times 5\ \mathrm{nm}^3$to $10 \times 10 \times 10\ \mathrm{nm}^3$. These sizes were selected based on the behavior of the average polarization magnitude, as shown in Fig. 3(e). For smaller cubes, with volumes below approximately 250 $\mathrm{nm}^3$, the average polarization magnitude is substantially reduced due to enhanced depolarization fields. Strong electrostatic fields at the interfaces favor the formation of vortex-like flux-closure configurations, in which the polarization rotates continuously and is significantly suppressed at the vortex core. The resulting vortex structures exhibit minimal variation along the vortex axis. In these simulations, we considered an $\epsilon_B = 45$ inside the material. These results are qualitatively consistent with previous reports [60–62]. However, the observed quantitative differences between the results obtained in this work and the literature highlight the need for standardized benchmark tests among microelectric solvers to ensure reproducibility and enable reliable cross-comparison of numerical implementations.

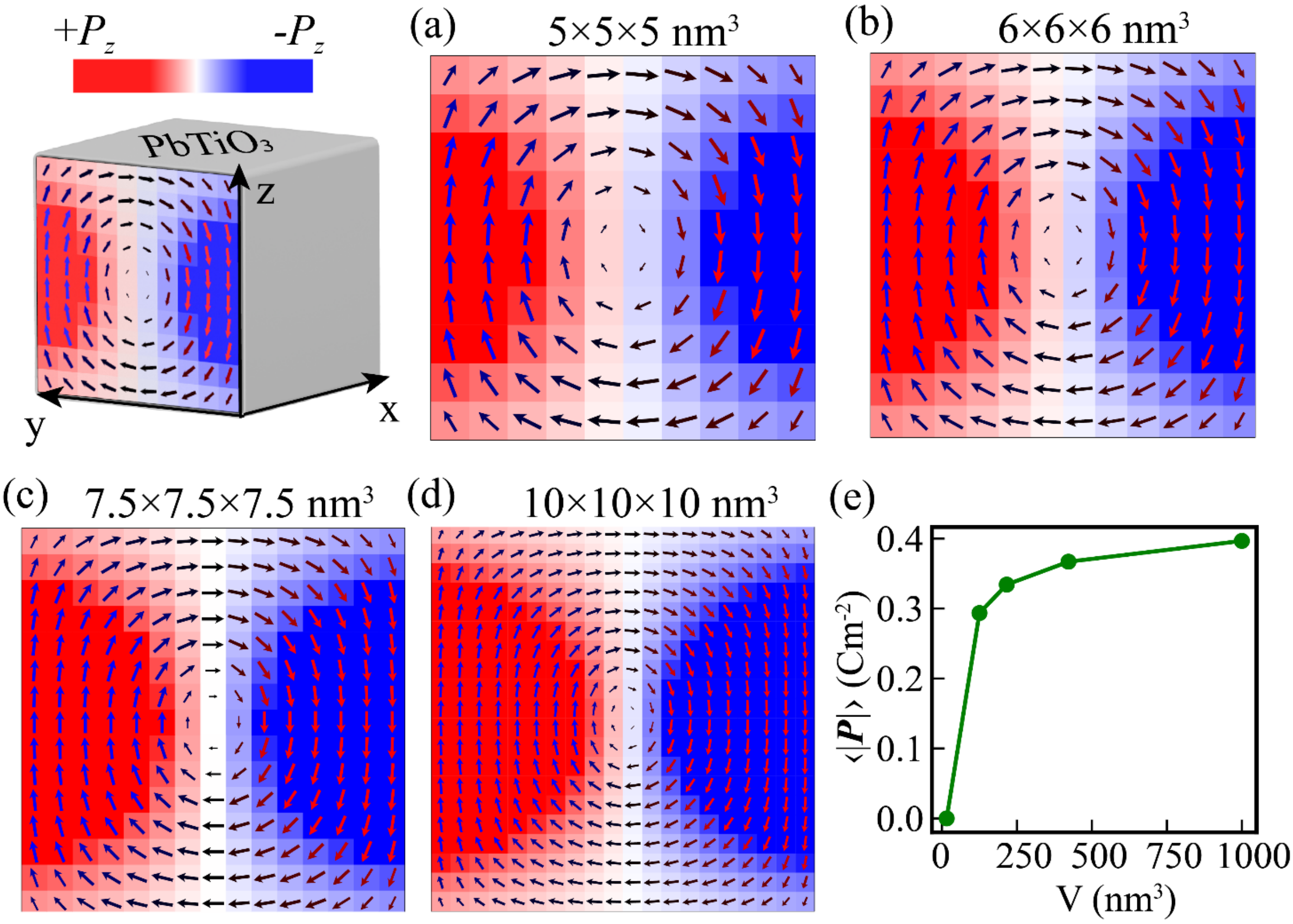

Figure 3. (a) scheme of the $PbTiO_3$ sample considered. (b)-(d) we show the cross section of the yz plane at the interface as a function of the cube size. (e) shows the average polarization magnitude as a function of volume.

### 4.4. Field based hysteresis loop in $PbTiO_3/SrTiO_3$

Ferroelectric materials can be used as non-volatile switchable components in devices such as memories or field-effect transistors because of their reversible spontaneous polarization [10]. Polarization switching is typically driven by alternating electric fields, whose frequencies strongly influence energy loss through ferroelectric hysteresis. We apply our solver to study frequency-dependent field-driven hysteresis under an alternating external electric field applied along the out-of-plane direction in $PbTiO_3/SrTiO_3$ bilayers. The Landau free energy parameters for $SrTiO_3$ were taken from [49], while the Landau and gradient energy parameters for $PbTiO_3$ were adopted from [48], as listed in Table 1 (parameters at $T = 300$ K and $\epsilon_B$ is treated as a fitting parameter and is discussed below). The choice of background permittivity is described below. The simulated sample size was 128 nm × 128 nm × 36 nm, consisting of a 20 nm thick $PbTiO_3$ layer sandwiched between 12 nm of $SrTiO_3$ and 4 nm of vacuum (see the sketch of the device as inset of Fig. 4(a)). The simulations are performed with a cubic cell size with lateral dimension Δ= 1.0 nm.

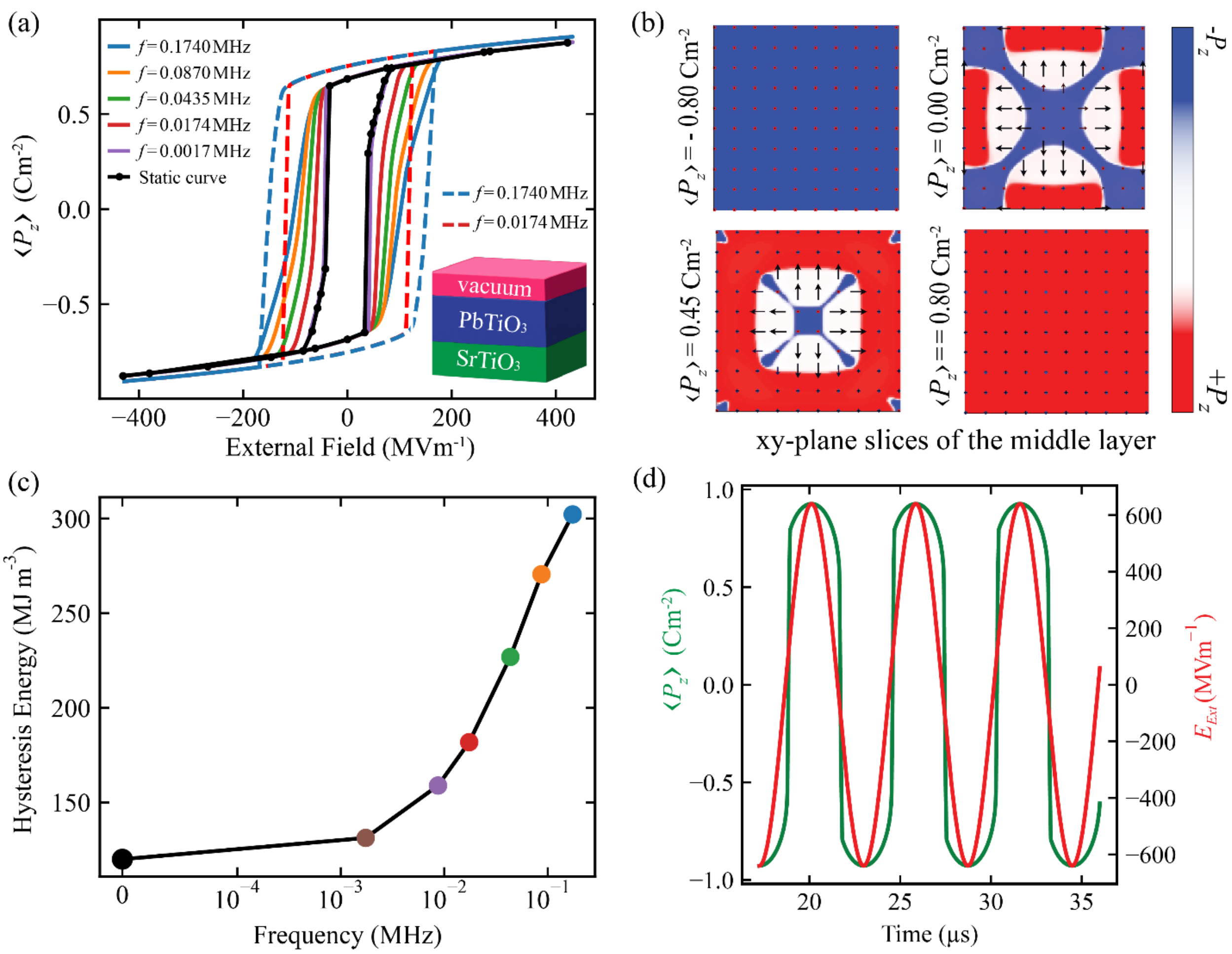

Figure 3. (a) (***P–E***) hysteresis loops of the $PbTiO_3/SrTiO_3$ bilayer calculated under various external field frequencies, together with the static loop obtained by sweeping a DC external field back and forth in the field region approximately -400 to 400 MV/m. (b) The domain structure showing switching from $-\langle P_z \rangle$ to $+\langle P_z \rangle$ for an AC-field frequency 0.0017 MHz. (c) Corresponding hysteresis energy as a function of the applied AC-field frequency; (d) Time evolution of the out-of-plane polarization $P_z$ and the applied external field $\boldsymbol{E}_{Ext}$ for the highest frequency considered 0.1740 MHz.

Figure 3(a) shows the ***P–E*** hysteresis loops of the $PbTiO_3/SrTiO_3$ for different frequency of the external applied field. As the sweep rate increases, the loops become noticeably wider, reflecting the growing phase lag between the electric field and the polarization response. To obtain the results with electrostatic field, within the finite sample consider, to reproduce the loop obtained in Ref. [49], we considered the approach similar to Ref. [42] to obtain $\epsilon_B = 649$, corresponding to the solid lines. The dashed lines show the loops in the absence of electrostatic field, revealing relatively higher coercive fields.

Overall, we notice that at high frequencies the polarization does not have sufficient time to fully relax to lower energy states or to complete domain stabilization under small applied fields, which leads to a larger coercive field and an overall increase in loop area. As the frequency is reduced, the loop width and coercive field consistently decrease; additionally, the change between consecutive low frequencies becomes progressively smaller. By the time the frequency reaches 0.0017 MHz, the loop nearly coincides with the static curve obtained from the DC field scan, indicating that the system has essentially reached its quasi-static loop. In this regime, the polarization evolves slowly enough to follow the field adiabatically, so further reductions in frequency do not significantly alter the hysteresis shape. The coercive field values extracted from these loops agree well with earlier reports for similar $PbTiO_3$-based structures [49].

The switching process can be seen more clearly in Figure 3(b) from the snapshots of the domain pattern in the central $xy$-plane of the $PbTiO_3$ associated with four values of the component of the average z-polarization as described below. We considered the lowest field frequency, which is sufficiently close to the quasi-static regime. Because the field changes very slowly, the system has enough time to develop complex intermediate domain structures during the reversal process. At $\langle P_z \rangle = -0.80\ \mathrm{C\,m^{-2}}$, the entire film is in a uniform single-domain state. When the polarization reaches $\langle P_z \rangle = 0.0\ \mathrm{C\,m^{-2}}$, a characteristic *flower-like* domain configuration is visible, with domains oriented along all six tetragonal directions. Positive $P_z$ domains emerge and grow to partially compensate the remaining negative $P_z$ domains, which shrink in size. This results in a vanishing spatially averaged out-of-plane polarization. As the external field continues to increase, the $+P_z$ domains grow by effectively moving the domain walls at the expense of the negative ones, ultimately leading to a complete switching into a uniformly polarized $+P_z$ state.

Figure 3(c) shows the area of the hysteresis loop as a function of the frequency of the sinusoidal electric field excitation. This hysteresis includes the contribution from the static and dynamic energy losses and can be attributed mainly to time delay response of the polarization to the external electric field as shown as example in Figure 3(d) for the frequency 0.1740 MHz. As expected, the energy loss decreases as the frequency decreases, since slower field variations reduce the dynamic contributions due to the nucleation and growth of domains. As the frequency is reduced, the dynamical hysteresis loops approach the static one showing a saturation in the hysteresis energy loss with a trend similar to that observed experimentally [68,69] and qualitative similar to the one observed in ferromagnetic materials.

### 4.4. Skyrmion stabilization via strain

Recently, ferroelectric skyrmions [7,70] have attracted significant attention due to their potential for next-generation logic and nonvolatile memory applications, stemming from their topological stability, nanoscale dimensions, and electrically switchable polarization textures [1,3,4,71]. Skyrmions correspond to whirl-like configurations of a vector field in which the core polarization is antiparallel to the surrounding background and is enclosed by a continuously rotating polarization characterized by an integer winding number [72].

We investigate polar skyrmions, using the proposed microelectric solver, in the model system of the field: $PbTiO_3/SrTiO_3$. To do this, we impose a realistic epitaxial constraint that forces an out-of-plane easy axis for the polarization of the ferroelectric layer. Specifically, we write the quadratic contribution to the free energy as $(T - T_c)\left(\alpha_0\left(P_x^2 + P_y^2\right) + r_s\, \alpha_0 P_z^2\right)$ where $r_s$ is a dimensionless strain parameter [73–75]. We considered a $PbTiO_3/SrTiO_3$ bilayer composed of a $40 \times 40 \times 32$ nm$^3$ $PbTiO_3$ layer deposited on top of a $40 \times 40 \times 8$ nm$^3$ $SrTiO_3$ layer, with the entire structure surrounded by vacuum. Figures 5(a)-(d) show representative cross sections of the $PbTiO_3$ layer for different values of the strain parameter $r_s$, with the cross-sectional planes indicated schematically in Fig. 5(e). In the absence of strain ($r_s = 1.0$), a vortex tube configuration is stabilized. Increasing $r_s$, thereby enhancing the preponderance of the out-of-plane polarization easy axis, leads first to multidomain states and eventually to the formation of a skyrmion in the bulk of the $PbTiO_3$ layer for $r_s = 1.8$. Figure 5(f) shows the total energy of the system as a function of the applied epitaxial strain. The observed energy reduction correlates with the formation of multidomain and skyrmionic configurations.

Several features characterize the stabilized skyrmion. At the interfaces, electrostatic effects strongly suppress the polarization component normal to the surface, resulting in a structure resembling cocoon-like textures reported in magnetic multilayers [76,77]. The interfacial electrostatic field also introduces a confining effect that influences the square-like shape of the skyrmion. Furthermore, the resulting configuration exhibits a Néel-type character, consistent with a flux closure and a 180° rotation of the polarization at the top and bottom interfaces, in agreement with observations in magnetic multilayer skyrmionic systems [78].

The Landau free-energy parameters for $PbTiO_3$ are summarized in Table I, while the gradient energy coefficients are taken from Ref. [36]: $G_{11} = 3.46 \times 10^{-10}\,\mathrm{Nm^4C^{-2}}$, $G_{12} = 0$, *and* $G_{44} = G'_{44} = G_{110} = 1.73 \times 10^{-10}\,\mathrm{Nm^4C^{-2}}$. Simulations were performed at 300 K using a uniform discretization of $0.5 \times 0.5 \times 0.5$ nm$^3$. For $r_s = 1.0$, the system was initialized from a random polarization configuration. For higher values of $r_s$, configurations were obtained adiabatically by using the relaxed solution at the preceding strain value as the initial state. These results illustrate the strain-controlled stabilization of skyrmions in ferroelectric layers.

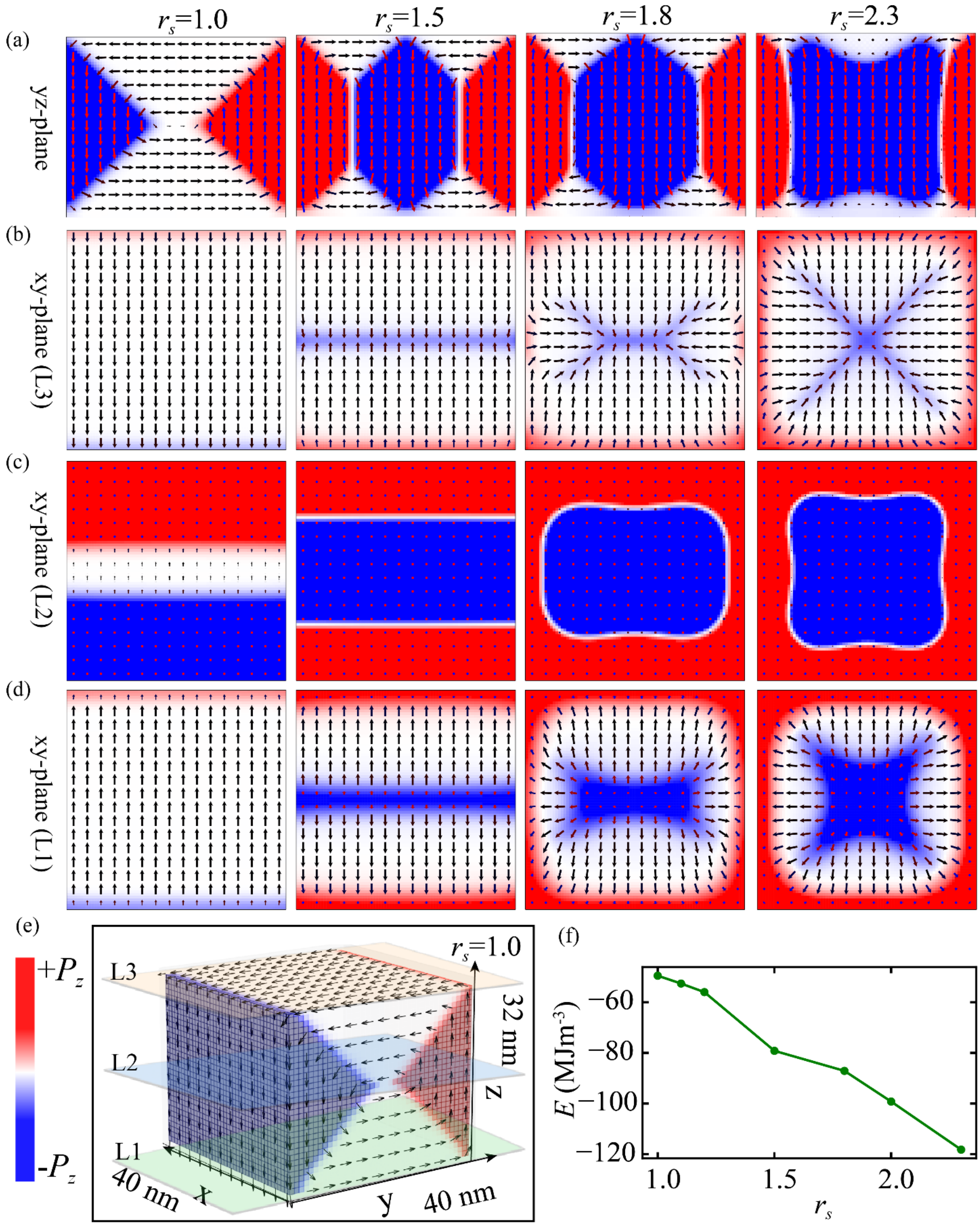

Figure 4. Strain-induced stabilization of a skyrmion in a $PbTiO_3$ sample. Panels (a)–(d) show cross-sectional views of the system under different values of the out-of-plane strain parameter. The color map represents the out-of-plane polarization component $P_z$, while arrows indicate the in-plane polarization components. The bottom panels display (e) a schematic of the sample highlighting the cross-sectional planes and (f) the average energy density as a function of the strain parameter $r_s$.

## 5. Conclusion

We have introduced PETASPIN_microelectrics, a GPU-accelerated phase-field solver for ferroelectric materials that incorporates a fully resolved electrostatic field. The solver addresses key limitations of existing CPU- and GPU-based implementations and enables accurate simulations of large ferroelectric systems while properly accounting for finite-size and boundary effects. Also of note is the fact that, despite massive advances in the past years [79–86], simulation methods based on machine-learning techniques continue to face challenges to address mesoscale phenomena (millions of atoms), particularly when the shape of the samples and/or the boundary conditions (e.g., in particular electrostatic) are non-trivial. In contrast, the present framework offers a robust and scalable physics-based approach valid across a wide range of materials, geometries, and field conditions [24].

The solver was systematically validated against analytical results for ferroelectric domain-wall profiles and reproduces Landau-based phase transitions in $BaTiO_3$ with machine-level numerical precision ($10^{-15}$). The model captures temperature-driven hysteresis behavior and sharp polarization changes at phase-transition boundaries. We further investigated field-driven, frequency-dependent hysteresis loops in $PbTiO_3/SrTiO_3$ bilayers, revealing reduced energy dissipation at low excitation frequencies. By simulating large samples of $PbTiO_3/SrTiO_3$ bilayers, we were also able to investigate the formation of Nèel electric skyrmions. By explicitly resolving both the polar order parameter and the associated long-range electrostatic field, including boundary effects, PETASPIN_microelectrics captures shape-induced effects that are essential for the stability of nontrivial polarization textures [65,66]. The resulting skyrmion configuration resembles cocoon-like states and skyrmion tubes reported in magnetic multilayers, where similar textures arise from the interplay between magnetostatic interactions and Dzyaloshinskii–Moriya interactions induced by inversion symmetry breaking [76,78]. This analogy opens a route for systematic comparison of topological textures in ferromagnetic and ferroelectric systems. For example, while ferroelectric hopfions have been shown to be stabilized by electrostatic interactions [65] magnetostatic interactions are generally considered detrimental to hopfion stability in magnetic materials [87,88], highlighting an important contrast that will be the focus of future investigations using PETASPIN_microelectrics.

The computational efficiency of PETASPIN_microelectrics enables large-scale simulations of finite ferroelectric systems and systematic exploration of broad parameter spaces, both of which are necessary to identify the conditions that stabilize complex topological textures. Beyond direct physical analysis, the solver also provides a robust platform for generating reliable datasets for data-driven methods, including machine-learning-assisted approaches. PETASPIN_microelectrics therefore enables accurate investigation of finite-size ferroelectric phenomena and provides a physically grounded framework for data generation, validation, and benchmarking of reduced-order models.

Overall, these results demonstrate the accuracy, scalability, and extensibility of PETASPIN_microelectrics as a platform for investigating ferroelectric microstructures and emergent topological states. Future developments will focus on incorporating strain and flexoelectric effects, as well as more general electrostatic and elastic boundary conditions, further extending the applicability of the solver to the modeling and design of nanoscale ferroic devices and topological architectures.

**Acknowledgements**

The authors thank John Mangeri for fruitful discussions on the benchmark problems. This research is part of the TOPOCOM project, which is funded by the European Union's Horizon Europe Programme Horizon.1.2 under the Marie Skłodowska-Curie Actions (MSCA), Grant Agreement No. 101119608.